\documentclass{article}
\usepackage{arxiv}
\usepackage{amsmath}
\usepackage{algorithmic}
\usepackage[ruled,vlined]{algorithm2e}
\usepackage{graphicx}
\usepackage{setspace}
\usepackage{textcomp}
\usepackage{balance}
\usepackage{multirow}
\usepackage{booktabs}
\usepackage{textcomp}
\usepackage{xcolor}
\usepackage{subfigure}
\usepackage{float}
\usepackage{stfloats}

\newcommand{\CASE}[1]{\STATE \textbf{Case} #1\textbf{:} \begin{ALC@g}}
\newcommand{\ENDCASE}{\end{ALC@g}}

\newcommand{\DEFAULT}{\STATE \textbf{Default:} \begin{ALC@g}}
\newcommand{\ENDDEFAULT}{\end{ALC@g}}
\newcommand{\DEFAULTLINE}[1]{\STATE \textbf{default:} }
\newcommand{\ie}{{\textit{i.e.}},\xspace}
\newcommand{\eg}{{\textit{e.g.}},\xspace}
\newcommand{\etc}{\textit{etc.}\xspace}

\usepackage{url}
\usepackage{hyperref}
\hypersetup{hidelinks}

\title{{\sc deGraphCS}: Embedding Variable-based Flow Graph for Neural Code Search}

\author{
  Chen Zeng \\
  School of Computer\\
  National University of Defense Technology\\
  \texttt{zengchen15@nudt.edu.cn} \\
  \And
  Yue Yu \\
  School of Computer\\
  National University of Defense Technology\\
  \texttt{yuyue@nudt.edu.cn} \\
  \And
  Shanshan Li \\
  School of Computer\\
  National University of Defense Technology\\
  \texttt{shanshanli@nudt.edu.cn} \\
  \And
  Xin Xia \\
  School of Computer\\
  Monash University\\
  \texttt{xin.xia@monash.edu} \\
  \And
  Zhiming Wang \\
  School of Computer\\
  National University of Defense Technology\\
  \texttt{wangzhiming14@nudt.edu.cn} \\
  \And
  Mingyang Geng \\
  School of Computer\\
  National University of Defense Technology\\
  \texttt{gengmingyang13@nudt.edu.cn} \\
  \And
  Linxiao Bai \\
  School of Computer\\
  National University of Defense Technology\\
  \texttt{linxiao\_b@nudt.edu.cn} \\
  \And
  Wei Dong \\
  School of Computer\\
  National University of Defense Technology\\
  \texttt{wdong@nudt.edu.cn} \\
  \And
  Xiangke Liao \\
  School of Computer\\
  National University of Defense Technology\\
  \texttt{xkliao@nudt.edu.cn} 
}

\begin{document}
\maketitle
\begin{abstract}
With the rapid increase in the amount of public code repositories, developers maintain a great desire to retrieve precise code snippets by using natural language.
Despite existing deep learning based approaches (\eg DeepCS and MMAN) have 
provided the end-to-end solutions (\ie accepts natural
language as queries and shows related code fragments retrieved directly from code corpus),
the accuracy of code search in the large-scale repositories is still limited by the code representation (\eg AST) and modeling (\eg directly fusing the features in the attention stage). 
In this paper, 
we propose a novel learnable \textit{de}ep \textit{G}raph for \textit{C}ode \textit{S}earch (called {\sc \textbf{deGraphCS}}), to transfer source code into variable-based flow graphs based on the intermediate representation technique, which can model code semantics more precisely compared to process the code as text directly or use the syntactic tree representation.
Furthermore, we propose a well-designed graph optimization mechanism to refine the code representation, and apply an improved gated graph neural network to model variable-based flow graphs. To evaluate the effectiveness of {\sc \textbf{deGraphCS}}, we collect a large-scale dataset from GitHub containing 41,152 code snippets written in C language, and reproduce several typical deep code search methods for comparison. Besides, we design a qualitative user study to verify the practical value of our approach. The experimental results have shown that {\sc \textbf{deGraphCS}} can achieve state-of-the-art performances, and accurately retrieve code snippets satisfying the needs of the users.


\end{abstract}


\keywords{Code search, graph neural networks, intermediate representation,  deep learning}

 \maketitle

\section{Introduction}
Code search has received increasing attention in recent years \cite{chen2018neural,gu2018deep,cambronero2019deep,shuai2020improving,wan2019multi}. The goal of code search is to retrieve code fragments which best meet developers’ need by performing natural language queries over a large code corpus. With the availability of immense and rapidly growing source code repositories such as GitHub and Stack Overflow, it is more convenient for developers to search the needed code with certain functionality and reuse it in their own programs. However, increasingly complex and diverse code implementations also bring great challenges to perform a precise code search.

In the early stage, code search approaches were proposed on the basis of information retrieval techniques, especially key words matching mechanism \cite{brandt2010example,campbell2017nlp2code,chan2012searching,holmes2009end,keivanloo2014spotting,li2016relationship,lv2015codehow,mcmillan2011exemplar,mcmillan2011portfolio}. 
However, a common problem of these works is the lack of structural or semantic information from the source code since they simply consider code and queries as plain texts. Recently, deep learning technologies have been applied to represent code and queries for code search \cite{chen2018neural,gu2018deep,cambronero2019deep,shuai2020improving,wan2019multi} to tackle the above issues. 
The typical approach, called DeepCS~\cite{gu2018deep}, presented a code search engine by learning a joint-embedding of a method description and its corresponding code snippet. 
Moreover, Wan et al.~\cite{wan2019multi} designed a Multi-Modal Attention Network (MMAN) to capture various code features simultaneously, such as code tokens, abstract syntactic tree (AST) and statement-based control-flow graph (S-CFG).

However, existing deep learning based approaches are still limited from two major aspects. First, in reality, code with different syntax may achieve the same functionality, while code with similar structural features may express totally different code semantics. Thus, the token (\eg method name or identifiers) and structural features (\eg AST or S-CFG) are hard to precisely express the in-depth semantics of source code in various forms (as shown in Fig. 1 and Fig. 2).
Second, existing methods cannot fully exploit multiple valuable features extracted from the source code. In concrete, some models did not fuse different modalities of source code effectively, which does not bring much improvement yet increases the complexity instead. For example, MMAN proposed an attention network to assign learnable weights to three different modality of source code (\ie Token + AST + S-CFG), while only outperforming the single token-based modality by 4.63\% and 6.12\% in terms of MRR (mean reciprocal rank) and SuccessRate@1.

These aforementioned limitations inspire us to design a model to effectively integrate several deep semantic information and learn a precise code representation. In our work, we explore a novel code representation based on data and control flow extracted from LLVM IR (Intermediate Representation)~\cite{lattner2004llvm}, one type of intermediate code acquired from source code. Different from the existing statement-based data and control flow representation method \cite{BenNun2018Nips}, we refine the construction of the variable-based flow graph to better describe the dependencies between code variables. In concrete, the nodes in the graph represent the tokens appeared in LLVM IR and the edges represent the data and control dependencies between the tokens. Furthermore, we design an optimization mechanism while modeling the graph to remove the redundant information brought by LLVM IR without changing the semantics. Finally, we employ an attentional gated graph neural network to embed the flow graph into a high-dimensional vector space to further perform code search tasks. Through this procedure, multiple semantic features of code, \ie tokens, variable-based data and control flow, can be simultaneously represented and accurately express the deep semantics of code.

To evaluate the effectiveness of our proposed model, we collect our dataset from GitHub containing 41,152 code snippets written in C language and perform code search experiments. Experimental results show that {\sc deGraphCS} improves the top-1 hit rate of code search from 34.05\% to 43.05\% when compared with the state-of-the-art methods. To simulate the actual code search scenario, we design an online code search tool, which takes 50 practical descriptions randomly chosen from the test set as candidate queries. For each query, 5 experienced participants manually label the relevant results they need returned by our proposed model {\sc deGraphCS} and three competitive approaches (\ie DeepCS, UNIF and MMAN). The results of automatic evaluation and manual evaluation both confirm the effectiveness of {\sc deGraphCS}.

The main contributions of this paper are summarized as follows:
\begin{itemize}
\item We propose a novel semantic code representation method called {\sc deGraphCS}, which can integrate token, data flow and control flow into a variable-based graph more precisely than traditional approaches (\eg AST). All of our code and data are released at \url{https://github.com/degraphcs/DeGraphCS}.
\item 
We design a graph optimization mechanism to streamline the representation of graph by reducing 51.88\% redundant nodes, which can significantly improve the performance of {\sc deGraphCS} by 13.77\% and 18.92\% in terms of MRR and SuccessRate@1.
\item We collect a large-scale dataset from GitHub containing 41,152 code snippets written in C language, and reproduce several competing code search models to make comparison.
\item We conduct experiments on the trained models, and the results of automatic evaluation demonstrate that {\sc deGraphCS} outperforms the state-of-the-art method (\ie MMAN) by 19.14\% and 26.43\% in terms of MRR and SuccessRate@1. Besides, {\sc deGraphCS} achieves the best performance in our qualitative user study.
\end{itemize}

The remainder of this paper is organized as follows. In Section \ref{motivate}, we present two motivating examples. In Section \ref{model}, we first provide an overview of our proposed model and then describe the details of each part in our model. In Section \ref{experiments}, we elaborate the experimental setup and report the experimental results. In Section \ref{related_work}, we briefly review the related works. Finally, in Section \ref{conculsion}, we conclude our study and future work.
\section{Motivating Example}

\label{motivate}
\begin{figure}
\label{fig_motivate}
    \centering
 \centering
		\includegraphics[width=0.99\textwidth]{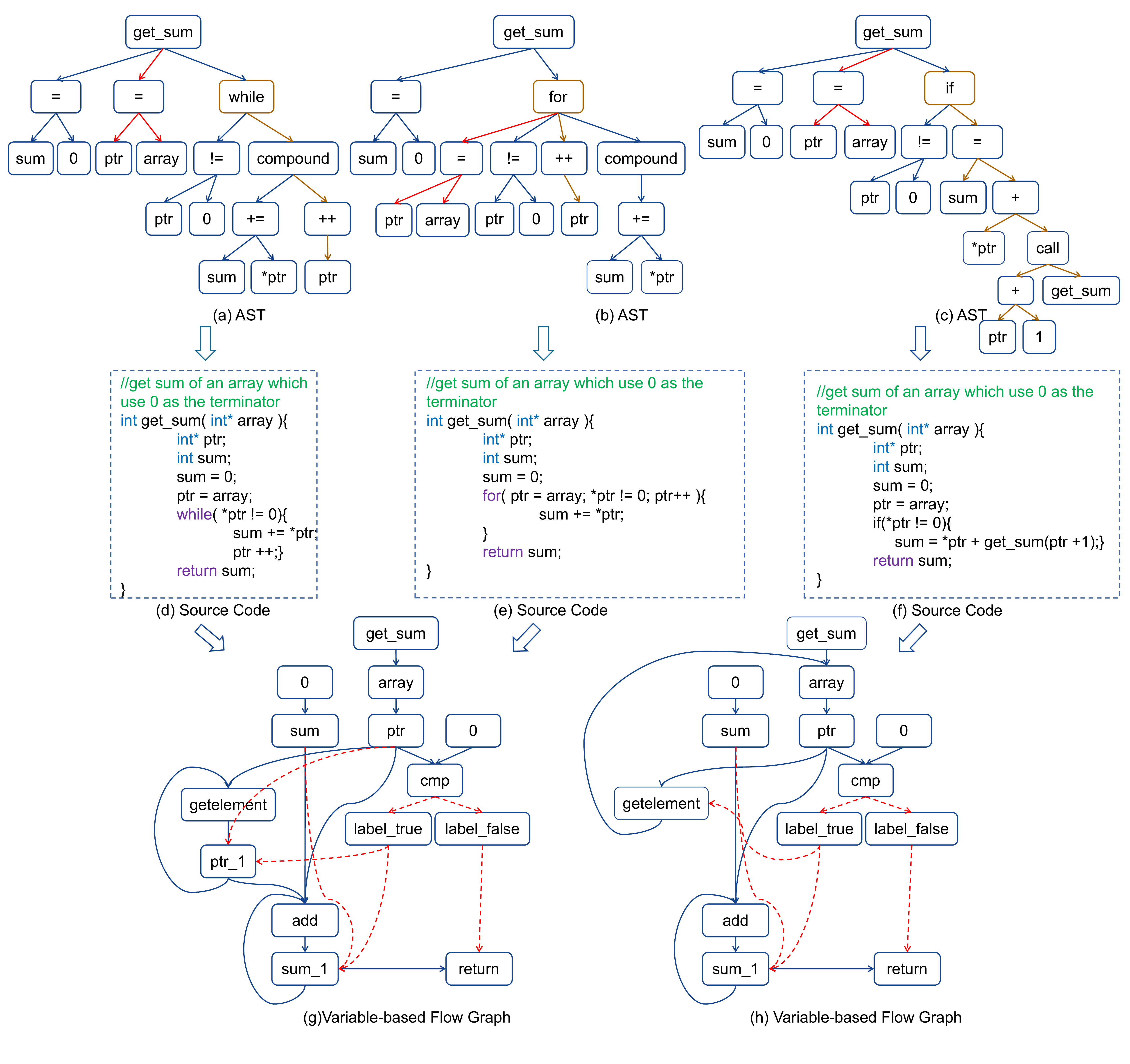}
	\caption{The first illustrative example shows the code snippets with the same semantics and their corresponding ASTs and variable-based flow graphs.}
\end{figure}

In this section, we show two motivating examples in Fig. 1 and Fig. 2 to illustrate the advantages of our semantic feature-based code representation approach over the existing methods which are based on structural features. Here, we argue that code snippets with same functionality may have different implementation, while code snippets with totally different code semantics may have similar structural features. Therefore, we need to precisely represent the source code on the semantics for performing a satisfying code search task. In other words, we need to find a more precise representation for source code, making semantically similar code have similar representation. 

Fig. 1(d-f) show three simple C code snippets as example, all of them aim to sum the values in the specified array. Fig. 1(a-c) represent the corresponding ASTs of the three code snippets. Our variable-based flow graph constructed from LLVM IR is illustrated in Fig. 1(g) and Fig. 1(h). Here, nodes in the flow graph are tokens of the intermediate code (\eg generated by LLVM IR) and the edges either represent data dependencies (shown in solid line of light blue) or control dependencies (shown in dotted line of red).
\begin{figure}
\label{fig_motivate2}
    	\centering
		\includegraphics[width=0.6\textwidth]{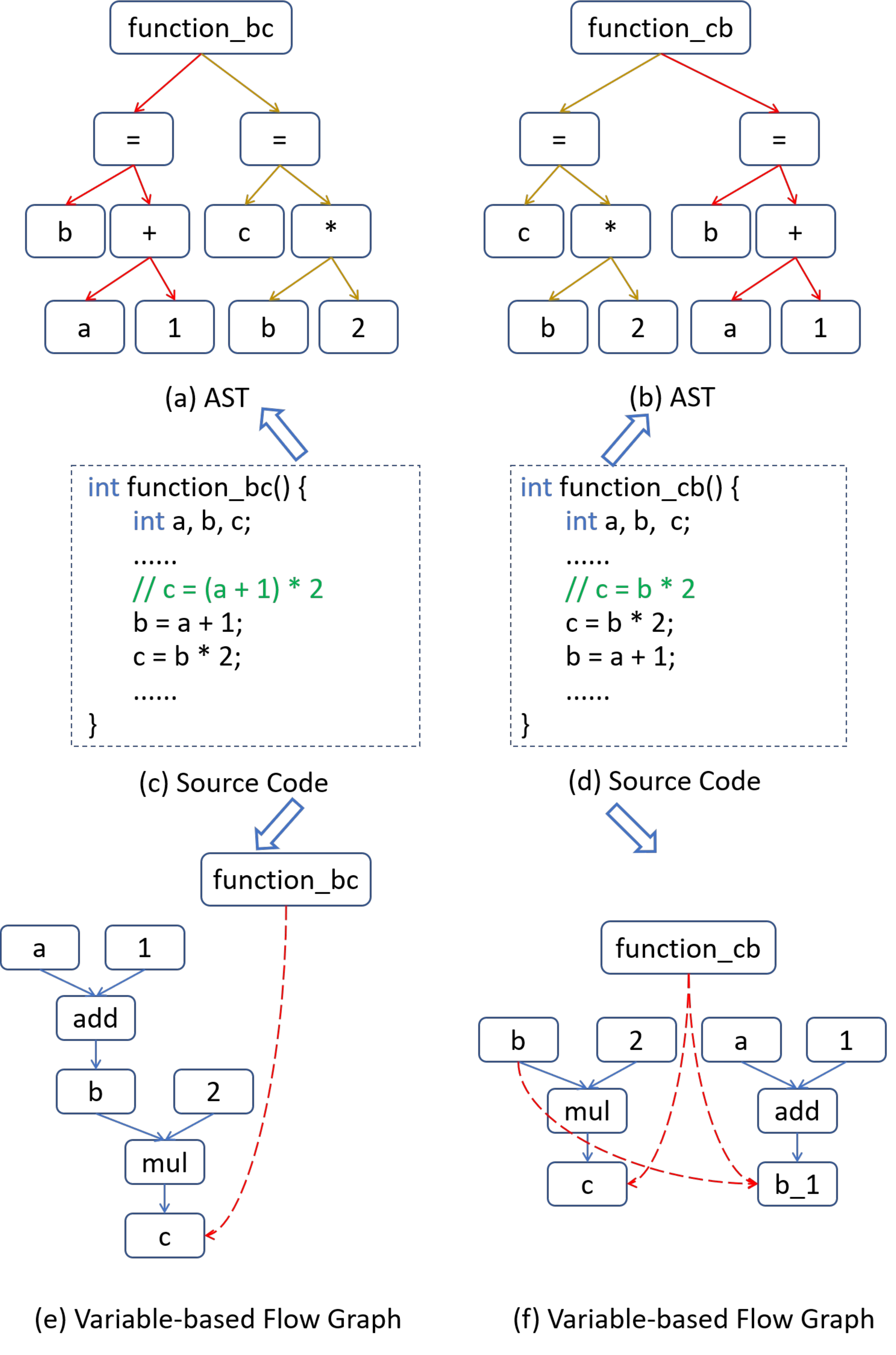}
	\caption{The second illustrative example shows the code snippets with different semantics and their corresponding ASTs and flow-based graphs.}
\end{figure}
From Fig. 1(d-f), we can clearly see that the three code examples implement the same function but have different writing format ( ``for'', ``while'' and recursive loop separately), which result in totally different syntactic structure in Fig. 1(a-c). The difference between the three ASTs is highlighted in red and yellow. In concrete, in Fig. 1(a) and Fig. 1(c), the parent node of the sub-tree connected in red line is ``get\_sum'', while in Fig. 1(b), the corresponding parent node is ``for'' instead. Moreover, in Fig. 1(c), ``if'' node's sub-tree connected in yellow line is completely different from the sub-tree of ``while'' node and ``for'' node in Fig. 1(a) and Fig. 1(b). However, as shown in Fig. 1(g-h), when exploiting our variable-based flow graph, the three code snippets with the same semantics are represented almost the same. In concrete, except for node ``ptr\_1'' and ``ptr'', every node in Fig. 1(g) can find corresponding node in Fig. 1(h). And compared with node in Fig. 1(g), the corresponding node have same control dependency and data dependency (\eg in Fig. 1(g) and Fig. 1(h), two ``return'' nodes both depend on ``sum\_1'' and ``label\_false''). In fact, in Fig. 1(g), both ``ptr\_1'' and ``ptr'' refer to variable ``ptr'', their dependencies are exactly the same as the ``ptr'' in  Fig. 1(h). In short, almost every data and control dependency in Fig. 1(g) can find corresponding one in Fig. 1(h). And the data and control dependency imply code semantics. Thus Fig. 1(g) and Fig. 1(h) can be regarded as same on code semantics. In all of the above, our flow graph establish an unifying representation for semantically similar code. And the unifying representation relies on the fact that data and control dependencies usually capture deeper code semantics than shallow syntactic features. Therefore, we argue that code snippets with similar semantics should maintain similar code representations.

Furthermore, to better explain our motivation that code snippets expressing different semantics should differ in code representations, we construct two segments of code snippets (shown in Fig. 2(c) and Fig. 2(d)) as the second example. Fig. 2(a)(b) and Fig. 2(e)(f) depict the AST and our variable-based flow graph of each code snippet. As shown in Fig. 2(c) and Fig. 2(d), although the two code examples consist of exactly the same statements, the different order will lead to different functionalities while running. From Fig. 2(a) and Fig. 2(b), we can see that the two tree structures are the same except for the relative positions of the two sub-trees of the root node (connected in red/yellow respectively), and the existing works like MMAN will finally establish the same representation when applying Tree-LSTM \cite{Kai2015Tree-LSTM}. However, the corresponding variable-based flow graphs differ significantly in Fig. 2(e) and Fig. 2(f) because of the underlying data dependencies between variables. The two examples have shown that our proposed variable-based flow graph representation can represent precise semantics of code while the syntactic structures fail.

Inspired by the above two examples, we conclude that data and control dependencies can complement the drawbacks of the structural features-based code representation methods. However, the existing works extract data and control flows on the basis of statements, obtaining vector representations of code snippets by applying word embedding technique such as skip-gram \cite{BenNun2018Nips}. A prominent problem is that coarse-grained statements usually cannot accurately capture the correlation between the tokens of code and query. Therefore, we propose our fine-grained variable-based flow graph method to precisely model the relationship between tokens in code snippets.

\section{The Proposed Model}
\label{model}
In this section, we first introduce an overview of our proposed network architecture. Then, we present our neural code representation mechanism, including background of compilation and LLVM IR, the variable-based flow graph building mechanism and optimization mechanism. Finally, we present our comment description representation and model learning mechanism in detail.

\subsection{An Overview}
\begin{figure*}[t]
\centering
\includegraphics[width=.99\linewidth]{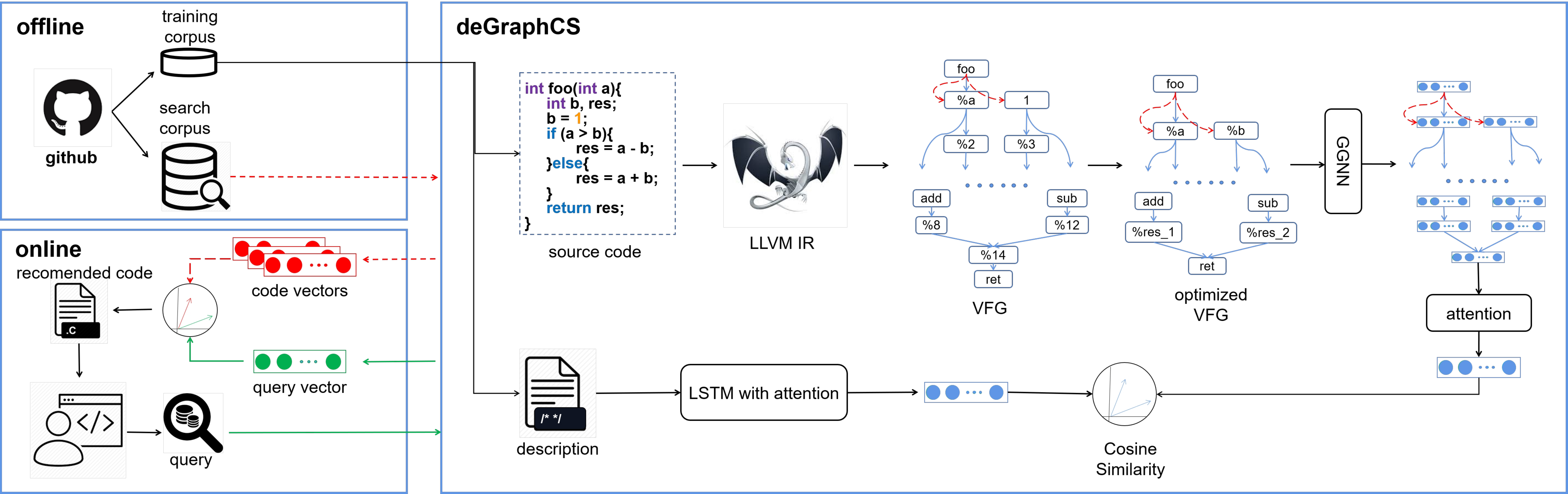}
\caption{The overall workflow of {\sc deGraphCS}, containing the offline data preparing, the online inference and the network architecture part.}
\label{fig1} 
\end{figure*}

Fig. 3 is an overview of workflow of {\sc deGraphCS} model, which is composed of three parts: the upper left part denotes the offline data preparing process; the bottom left part denotes the online inference process; the right part denotes the details of the network architecture. For the network architecture, {\sc deGraphCS} first embeds the neural code and comment to the vector representations, and then learns the relationship by minimizing the ranking loss function in the training process. We will describe each part of the architecture in the following sections.


\subsection{Neural Code Representation}

For code representation, we first integrate data dependencies and control dependencies into graphs by analyzing different kinds of LLVM IR instructions. In concrete, we construct the data dependencies based on the address operation instructions (\eg ``load'', ``store'') and the computation related constructions (\eg ``add'' and ``sub''). Besides, the control dependencies are constructed based on the jump instructions (\eg ``br'') and address operation instructions. The goal of code search is to better match the semantics in code with the keywords in queries, excessive information may hinder the model from learning the fine-grained relationship between the source code and queries. Therefore, we propose several mechanisms to optimize the graph with the aim of decreasing the noises in the model training process and improving the training efficiency. Finally, we feed the graph into a GGNN\cite{li2015gated} with attention mechanism to learn the vector representation of the code. In order to better explain our proposed neural code representation method, an illustrative example code associated with its IR, and our variable-based flow graph building and optimizing result is shown in Fig. 4.

\begin{figure*}[t]
\centering
\includegraphics[width=0.99\linewidth]{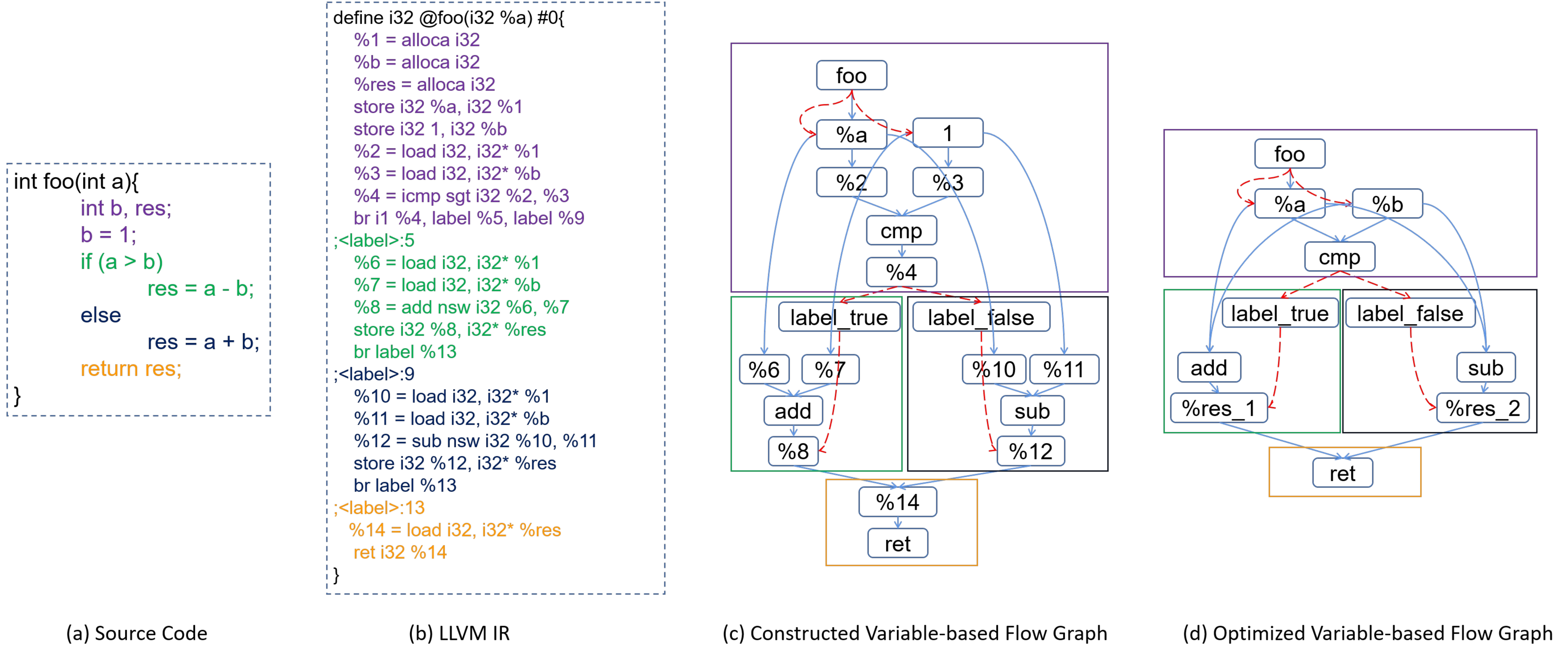}
\caption{An illustrative example shows a code snippet with its LLVM IR equivalent and our graph building as well as optimizing manner.}
\label{fig1} 
\end{figure*}

\subsubsection{Compilation and LLVM IR}

Most popular compilers, such as LLVM and GCC, support multiple programming languages and hardware targets. With the aim of avoiding duplications in code optimization techniques, the compilers require a strict separation among the source language, IR, and the target machine code which will be mapped to a specific hardware. LLVM IR supports various architectures and can represent optimized code inherently. The IR in LLVM is given in Static Single Assignment (SSA) form \cite{Cytron1991SSA}, which guarantees that every variable is assigned only once. As shown in Fig. 4(b), LLVM divides the IR statements into several blocks represented by the corresponding labels shown in different colors. For the instructions, regarding the third line in ``label 9'', an instruction (IR statement) in our algorithm is mainly composed of three parts: opcode (``sub''), operand ($\%10$, $\%11$) and result ($\%12$).


\begin{algorithm}
	\caption{Variable-based Flow Graph Building Process}
	\label{alg1}
	\begin{algorithmic}[1]
		\REQUIRE LLVM IR (shown in Fig. 4(b)). 
		\ENSURE the constructed variable-based flow graph (shown in Fig. 4(c)).
    \FOR{read the instructions of IR by line}
    \CASE{Computation instructions}
    \CASE{``call/invoke''}
    \STATE Build an edge parameters$\longrightarrow$ function name.
    \ENDCASE
    \DEFAULT
    \STATE Build an edge operands $\longrightarrow$ opcode.
    \STATE Build an edge opcode$\longrightarrow$ result.
    \ENDDEFAULT
    \ENDCASE
    \CASE{Address instructions}
    \FOR{each load instruction operated on $addr$ (\ie a = load addr)}
    \STATE value\_list = SearchCFG($addr$, $inst$).
    \STATE Build edges value\_list $\longrightarrow$ $a$.
    \ENDFOR
    \FOR{all instructions ``store $x$ addr''}
    \STATE Connect $x$ sequentially according to the order in CFG. 
    \ENDFOR
    
    \ENDCASE
    \CASE{``br'' (\ie br $\%val$, label 1,label2)}
    \STATE Build edges condition $\longrightarrow$ labels. 
    \STATE Build edges labels $\longrightarrow$  ``store'' variables.
    \ENDCASE
    \ENDFOR
	\end{algorithmic}
\end{algorithm}

\subsubsection{Building Variable-based Flow Graph (VFG)}

To derive a precise semantic representation of source code, we construct the graph at the granularity of variables to capture token information of source code. In concrete, we build data dependencies and control dependencies between variables to capture the data  and control flow information from source code. An illustrative example of the variable-based flow graph is shown in Fig. 4(c), which is the initial graph constructed from the LLVM IR shown in Fig. 4(b). The nodes in our graph can be variables, opcode or label identifiers, appearing in the figure as rectangles. Correspondingly, an edge either represents data dependency (in blue solid line), or control dependency (in red dotted line). Given the LLVM IR, the whole graph building process is recorded in \textbf{Algorithm 1}. In concrete, we first extract the identifiers in each IR instruction as nodes. Then, we build data dependencies and control dependencies between nodes according to different types of instructions as follows.

\noindent\textbf{{Data dependency.}} There exist data dependencies in the computation related constructions (\eg ``add'', ``sub'') and the address operation instructions (\eg ``load'', ``store''). First, we build data dependencies according to instructions which are related to computation such as ``add'', ``sub'' etc. For example, regarding the third line in the second block shown in green of Fig. 4(b) (``$\%8$ = add nsw i32 $\%6$, $\%7$''), we build data dependencies by linking the operands ($\%6$ and $\%7$) to the opcode (``add''), and then linking the opcode (``add'') to the result ($\%8$) shown in the green square of Fig. 4(c). We specially deal with the ``call/invoke'' instruction since the operands are the parameters of corresponding function. In concrete, when we build graph for current function (\eg ``get\_sum''), we treat the function call instruction in two situation. First, if the called function is the external function which is not the current function (``get\_sum''), we treat the name as opcode instead of ``call/invoke'', and link the operands to the called function name. Second, if the called function is the current function, it is regarded as a recursive call. Thus we regard the node which link to ``return'' node as the result of called function, and we link the result to the nodes which use the called function. Moreover, we regard the input of called function is passed into the parameter of current function, thus we link the input node to the parameter node. For example, when we construct data dependency for the function ``get\_sum(array)'' in Fig. 1(f), the function ``get\_sum(array)'' call itself recursively (``sum = *ptr + get\_sum(ptr +1);''). Thus we deal with the function call ``get\_sum(ptr +1)'' according to the second situation. As Fig. 1(h), we can find ``sum\_1'' is linked to ``return'' which implies ``get\_sum'' return ``sum\_1'' as result. Thus we link ``sum\_1'' to ``add'' which use the result of ``get\_sum'' to compute. Then we link ``getelement'' which denotes the input ``ptr +1'' to parameter node ``array'' of function ``get\_sum(array)''.

\begin{algorithm}
	\caption{Search All the Variables from $addr$ in $inst$}
	\label{alg1}
	\begin{algorithmic}[1]
		\REQUIRE $addr$, $inst$. 
		\ENSURE value\_list.
    \STATE  pre\_list $\leftarrow$ the last instructions pointed to inst.
    \FOR{each instruction pre\_inst in the pre\_list}
    \IF{pre\_inst has been searched}
    \STATE continue
    \ELSIF{pre\_inst is ``store $x$ addr''}
    \STATE Append the variable $x$ to the value\_list.
    \ELSE
    \STATE list = SearchCFG(addr, inst)
    \STATE Append all the values in list to value\_list
    \ENDIF
    \ENDFOR
	\end{algorithmic}
\end{algorithm}

Second, we build the data dependencies of variables in ``load'' and ``store'' instructions. In concrete, when the variables need to be used, LLVM will load the corresponding values from the allocated address by ``load'' instruction. Similarly, when variables need to be assigned new values, IR will store the new values into address by ``store'' instruction.
The reason to treat ``load'' and ``store'' instructions separately is that in these two instructions, the address may store multiple values from different variables, it is hard to build one-to-one mapping relationship between each address and variable. Therefore, we build data dependencies only between variables. For example, regarding the ``load'' instructions of the second block shown in green of Fig. 4(b), although the variables $\%6$ and $\%7$ are loaded from the address $\%1$ and $\%b$, the edges are connected from variables $\%a$ and $1$, which are the true sources. To achieve this, we need to traverse all the ``load'' instructions, and handle each instruction following the function in \textbf{Algorithm 2}.  

\noindent\textbf{{Control dependency.}} There exist control dependencies in the jump instructions (\eg ``br'') and the address operation instructions (\eg ``load'', ``store''). We exploit address operation instructions since multiple variables stored in an address usually maintain a sequential order. Thus, we complete control dependencies of variables through two aspects as follows:

First, we build control dependencies between variables and the condition identifiers. These conditions appear in condition jump instructions such as ``br'' instruction as shown in the last purple line of Fig. 4(b). Label identifiers are the entries of basic blocks, and the condition determines which label to jump. Thus, we construct control dependencies by linking the condition to all label identifiers. After that, to make the whole graph connected, we should build control dependencies between the label and the corresponding basic block's variables. In our algorithm, we link the label identifier to the variables in ``store'' instruction as shown in the red line from label\_true to $\%8$ of Fig. 4(c). The reason is that if a variable updates its value, a new value will be stored in the corresponding address by a store instruction.  

Second, multiple assignments of the same variable will generate different variables to be stored in the address and the variables usually maintain a sequential order. Therefore, we build control dependencies between these variables of the same address. As shown in line 13 of Algorithm 1, we need to traverse every store instruction which stores a variable to address \textbf{x}, and find the next ``store'' instruction which stores a new variable to address \textbf{x}, then build a connection from the previous variable to the latter variable until all the variables are connected in sequential order.

\subsubsection{Optimizing Variable-based Flow Graph}

After constructing the variable-based flow graph, we need to optimize the graph to decrease the noises and improve the training efficiency. The optimizing method is composed of the following four steps.

\begin{enumerate}
    \item First, in LLVM IR, variables in ``store'' instruction are named with numbers (\eg $\%1, \%2$). Since the goal of code search is to better match the tokens (\eg identifiers, functions names) in code with the keywords in queries, excessive numbers may hinder the model from filtering the critical information of the flow graph. Therefore, we replace the numbers with the corresponding variable names.
    
    \item Second, plenty of opcode are too trivial to represent the code semantics needed for code search. Therefore, we remove these opcode nodes from the graph by linking the predecessors of the opcode to their successors. In concrete, the trivial opcode can mainly be divided into three kinds. The first kind of opcode are those related to memory access and addressing since they operate on the variable address. The second are opcode related to conversion since they aim to transform the type of the data (\eg change ``int'' type to ``string'' type). The third are opcode related to operations on exceptions.
    
    \item Third, in LLVM IR, many temporary registers are generated to store the intermediate values and these registers have no corresponding variables. Thus, we remove these variable nodes by linking their predecessors to their successors.
    
    \item Fourth, we compress the control flow graph by merging the ``isolated'' blocks. In our constructed variable-based flow graph, suppose block $a$ is the predecessor of block $b$, if block $a$ has only one successor and block $b$ has only one predecessor, then we merge the two blocks to compress the control flow.
    
\end{enumerate}

\subsubsection{Graph2Vec}
Since our constructed graph is a directed graph with multiple types of edges, we utilize an improved gated graph neural network (GGNN) with attention mechanism to learn the vector representation of the code. GGNN is a neural network architecture for embedding graphs with multiple types of edges. In our graph $G\ =\ (V,E)$, $V$ denotes a set of nodes$(v,l_v)$, and $E$ denotes a set of edges$(v_i,v_j,l_{(v_i,v_j)})$. $l_v$ denotes the label of node $v$ which is consisted of the variables in IR instructions. $l_{(v_i, v_j)}$ denotes the label of the edge from $v_i$ to $v_j$ which includes two types: data dependency and control dependency. 

GGNN learns the vector representation of $G$ by message passing mechanism as follows. First, we initialize each node $v\in V$ with a one-hot embedding vector($h_v^0$) according to $l_v$. Then, we train the embeddings of all nodes through multiple iterations. In iteration $t$, each node $v_i$ obtains message $m_t^{v_j\mapsto v_i}$ from neighbour $v_j$ as: $m_t^{v_j\mapsto v_i}\ =\ W_{l_{(v_i,v_j)}}h_{v_j}^{t-1}$, 
$W_{l_{(v_i,v_j)}}$ is the weight matrix specified by the type of edges, it map message from neighbour $v_j$ into a shared space. The weight matrix is learned during training process. Then all message from the neighbours of $v_i$ are aggregated in the following equation:
\begin{equation}
    m_t^i\ =\ \underset{v_j\in Neibour(v_i)}{\Sigma}(m_t^{v_j\mapsto v_i})
\end{equation}
Then, GGNN uses GRU (Gated Recurrent Unit)\cite{Junyoung2014GRU} to update the embedding of each node $v_i$. GRU uses aggregated message and past state $h_{v_i}^{t-1}$ to update current state as  $h_{v_i}^{t}\ =\ GRU(m_t^i,h_{v_i}^{t-1})$. 
Finally, since different nodes contribute differently to the code semantics, we exploit the attention mechanism to calculate the importance of different nodes. We first allocate weights for each node $v_i$ as:
\begin{equation}\alpha_i\ =\operatorname{ sigmoid}(f(h_{v_i}) \cdot u_{vfg})\end{equation}
$\alpha_i$ denotes the weight of node $v_i$, $f(\cdot)$ denotes the linear layer, $\cdot$ denotes the inner project function and $u_{vfg}$ denotes the context vector which is a high level representation of the whole nodes in graph. $u_{vfg}$ is realized as a linear layer which randomly initialized and jointly learned during training. Then, we obtain the embedding of the whole graph $h_{vfg}$ as:
\begin{equation}
   h_{vfg} = \underset{v_i\in V}{\Sigma}(\alpha_i h_{v_i}).
\end{equation}

\subsection{Comment Description Representation}
For comment representation, we apply LSTM\cite{Hochreiter1997LSTM} to learn the corresponding representations. The embedding $h_{i}^{des}$ of each word in comment is calculated as $h_{i}^{des}\ =\ LSTM(h_{i-1}^{des},w(d_{i}))$,
where $i = 1,...,|d|$, $|d|$ denotes the length of the comment description, and $w$ denotes the word embedding layer to embed each word into a vector. Since different parts of the comment make different contributions to the final vector representation, we adopt an attention mechanism \cite{bahdanau2014neural} to capture the fine-grained relevance between the hidden states and the final comment representation. In concrete, we apply an attention layer to calculate the attention score $\alpha^{des}(i)$:
\begin{equation}
    \alpha^{des}(i)=\frac{\exp \left(f({h}_{i}^{des}) \cdot u^{des}\right)}{\sum_{k=1}^{n} \exp \left(f({h}_{k}^{des}) \cdot u^{des}\right)}
\end{equation}
where $\cdot$ denotes the inner project of ${h}_{i}^{des}$ and $u^{des}$, $f(\cdot)$ denotes a linear layer and $u^{des}$ denotes the context vector which is a high level representation of the whole tokens in comment. The context vector $u^{des}$ is randomly initialized and jointly learned during training. Then, the final representation of the comment description ${E}_{|d|}^{des}$ can be calculated as:
\begin{equation}
    {E}_{|d|}^{des}=\sum_{i=1}^{|d|} \alpha^{des}(i) {h}_{i}^{des}
\end{equation}

\subsection{Model Training}

Now we obtain all code representation ($C$) and description representation ($D$). To search the code precisely for each query, the model should make the code representation similar with the correct description representation, and make the code representation different with the incorrect description representation. In other words, for a code snippet representation $c\in C$, the associated correct description $d^+\in D$ and a randomly chosen incorrect description $d^-\in D$, we need to make the vector representation of the pair  $<c,d^+>$ similar and the vector representation of the pair $<c,d^->$ different. Therefore, we train the model by minimizing the loss function $L(\theta)$ in formulation of:

\begin{equation}
   L(\theta)\ =\ \underset{c\in C\ d^+,d^-\in D}{\Sigma}max(0,\beta -cos(c,d^+)+cos(c,d^-))
\end{equation}
where $\theta$ denotes the model parameters, $d^+$ denotes the correct description representation, $d^-$ denotes the incorrect description representation. The $cos(,)$ function is to measure the consine similarity between two vector representations, and $\beta$ denotes the constant margin.

\section{Experiments and Results}
\label{experiments}
In this section, to evaluate the performance of {\sc deGraphCS} in code search, we perform several experiments to answer the following research questions: 

\begin{itemize}
\item RQ1: How effective is our proposed {\sc deGraphCS}?
\item RQ2: What is the effectiveness of our natural integration of multiple information, \ie code tokens, variable-based data and control flow extracted from IR in {\sc deGraphCS} when compared with the existing attention-based multi-modal representation method?
\item RQ3: How does our graph optimizing mechanism affect the final retrieval performance?
\item RQ4: What is the performance of {\sc deGraphCS} when varying the comment length, code length, VFG nodes number and the longest path length of VFG?
\item RQ5: What is the performance of {\sc deGraphCS} for helping developers in actual programming?
\end{itemize}

RQ1 is to investigate whether {\sc deGraphCS} outperforms the state-of-the-art deep code search models. RQ2 is to evaluate the effectiveness of integrating different modalities of source code used in our work. RQ3 aims to investigate how our graph optimizing mechanism improves the training results. RQ4 is to test and verify the robustness of our proposed model when varying the comment length, code length, VFG nodes number and the longest path length of VFG. RQ5 aims to evaluate the performance of {\sc deGraphCS} compared with the state-of-the-art models in manual evaluation. 

\subsection{Experimental Setup}
Here, we first describe our experimental dataset and present three widely used evaluation metrics. Then we describe the details of the implementation and introduce the baseline models for comparison.

\subsubsection{Data Collection}
As described in Section \ref{model}, our {\sc deGraphCS} model needs a large-scale training corpus which contains sufficient code fragments and the corresponding comments descriptions. Unfortunately, we cannot get access to the datasets collected by the existing works like MMAN and UNIF. In fact, we have also considered the dataset released by DeepCS, while this dataset only contains the cleaned Java code. It is hard for us to generate the LLVM IR without raw data. Therefore, we re-construct a corpus of C code snippet, which are crawled from GitHub, to verify the performance of our proposed model. We choose C code snippets because C language is popular and LLVM IR has been widely used on C language.

To build the required dataset, we collect high-star C projects from GitHub (a popular open source projects hosting platform). Then we collect our dataset by selecting the C methods which contain the corresponding comment descriptions and can be compiled into LLVM IR from projects. For each C method $c$, we treat the first sentence appeared in the comment as the corresponding natural language query $q$ since it typically describes the functionality implemented by the method \cite{codesearchchanllege2020,hu2018deep}. To reduce bad comments as much as possible, we use regular expression to delete comments which are not related to method function. Furthermore, we filter the ($q$, $c$) pairs by the following rules:

\begin{itemize}
\item ($q$, $c$) will be filtered out if the code snippet $c$ is a constructor or a test method.
\item ($q$, $c$) will be filtered out if the length of the query $q$ is less than 3 words or longer than 30 words.
\item ($q$, $c$) will be filtered out if the length of the code snippet $c$ is less than 5 lines or longer than 30 lines.
\item If a ($q$, $c$) pair appears multiple times in the dataset, we will remove the duplication.
\end{itemize}

After collecting the corpus of commented code snippets, we then extract LLVM IR for our proposed model and other features of the source code for the baseline models, \ie method name, tokens, AST and  S-CFG. Finally, we obtain 41,152 samples which are more than the 28527 samples in MMAN\cite{wan2019multi}. And since the 41,152 samples are distributed in 1554 open source projects, it is general to train our model. Following \cite{wan2019multi}, we shuffle our dataset and split it into training set with 39,152 pairs and test set with 2,000 pairs respectively. In order to avoid the bias resulting from evaluating on isolated dataset, we resort to the automatic evaluation metrics on corpus with the ground truth. 

Furthermore, for performing manual evaluation, we first randomly select 100 descriptions from the test set, then we carefully choose 50 descriptions which are easiest to understand. And we rewrite the 50 descriptions (\eg add some conjunction) as test queries to ensure that they are similar enough to the real-world user queries. We construct a search codebase containing 30799 C code snippets without the training samples to guarantee the fairness. 5 experienced participants are required to select the relevant results returned by each model and record the score.

\subsubsection{Evaluation Metrics}

We choose two common metrics to measure the performance of code search: SuccessRate@k, and Mean Reciprocal Rank (MRR). 

For the automatic and manual evaluation, we adopt both SuccessRate@k and MRR to assess the performance of a code search model with respect to a set of queries. SuccessRate@k represents the percentage of queries for which more than one correct snippet succeed to exist in the top $k$ ranked snippets returned by a search model, which is calculated as: 
$\text {SuccessRate } @ k=\left(\frac{1}{|Q|} \sum_{q=1}^{Q} \delta\left(\text {Rank }_{q} \leq k\right)\right)$,
where $Q$ denotes the set of queries in our automatic evaluation; $Rank_{q}$ denotes the highest rank of the hit snippets in the returned snippet list for the query; $\delta()$ denotes a indicator function that returns 1 if the Rank of the $q_{th}$ query ($Rank_{q}$) is smaller than $k$ otherwise returns 0. SuccessRate@k is important because a better code search engine should allow developers to find the desired snippet by inspecting fewer results. Following MMAN, we evaluate SuccessRate@1, SuccessRate@5, SuccessRate@10 respectively and a higher SuccessRate@k value implies a better performance of the code search model.

We also use MRR to measure the ranking of the search results of each model. MRR is the average of the reciprocal ranks of all queries $Q$. The reciprocal ranks is the inverse of the highest rank of hit code, \ie Rank. The computation of MRR is:
$ \mathrm{MRR}=\frac{1}{|Q|} \sum_{q=1}^{|Q|} \frac{1}{\mathrm{Rank}_{q}}$,
where $Q$ denotes the set of queries in the automatic evaluation; $Rank_{q}$ denotes the rank of the ground-truth code corresponded to the the $q_{th}$ query. The higher the MRR value, the better the code search performance is.

\subsubsection{Baseline Models}
We compare the effectiveness of {\sc deGraphCS} with 3 state-of-the-art deep learning-based code search methods:

\begin{itemize}
\item \textbf{DeepCS}: DeepCS~\cite{gu2018deep} is a deep code search engine using deep neural networks. Instead of matching text similarities like traditional works, DeepCS learns a unified vector representation of both source code and the corresponding natural language query. We use the official implementation provided by the authors \footnote{\url{https://github.com/guxd/deep-code-search}}.
\item \textbf{UNIF}: UNIF \cite{cambronero2019deep} is a supervised extension of the base NCS technique \cite{sachdev2018retrieval}. UNIF maintains significantly lower complexity than previous sequence-of-words-based networks by using a bag-of-words-based network.
\item \textbf{MMAN}: MMAN is a novel multi-modal neural network for code search \cite{wan2019multi}. MMAN proposes an attention mechanism to incorporate multiple features including code tokens, AST and S-CFG to learn a more comprehensive representation for code understanding.
\end{itemize}

Moreover, to answer RQ2 (how effective is our graph-based integration compared with multi-modal attention) and RQ3 (how graph optimization in {\sc deGraphCS} affects its effectiveness), we compare {\sc deGraphCS} with some of its variants as follows:
\begin{itemize}
\item \textbf{MMAN (Token+V-DFG+V-CFG)}: In this variant, we exploit the features of {\sc deGraphCS}, \ie tokens of code, variable-based data and control flow graph and fuses them in a multi-modal neural network with attention mechanism. In other words, the features used in MMAN are replaced with the features used in {\sc deGraphCS}. This variant is used to validate that our graph building and optimizing mechanism is more effective than the previous multi-modal incorporation mechanism.
\item \textbf{{\sc deGraphCS}-noGO}: In this variant, we remove the graph optimization mechanism from {\sc deGraphCS}. This variant is used to validate that it is necessary to remove the redundant information in the initial constructed graph and verify how this mechanism can improve the training results.
\end{itemize}

\subsubsection{Implementation Details}
To train our proposed model, we first shuffle the training data and set the mini-batch size to 16. We build two separate vocabularies for comments and LLVM IR tokens and limit their size of vocabulary to 10,000 and 15,000 respectively to store the most frequently appeared tokens in the training dataset. For each batch, comments are padded with a special token ``PAD'' to the maximum length which is set to 30 in our experiments. All tokens in our dataset are converted to lower case and parsed into a sequence of tokens according to camel case and ``\_'' if exists. We set the word embedding size to 300. For LSTM and GGNN unit, we set the hidden size to 512. Besides, we set 5 rounds of iteration for GGNN. The margin is set to 0.6. We update the parameters via AdamW optimizer \cite{Ilya2017AdamW} with the learning rate 0.0003. All the models in this paper are trained for 200 epochs. All the experiments are implemented using Pytorch 1.6 framework with Python 3.6, and the experiments are conducted on a server with one Nvidia Tesla V100 GPU, running on Ubuntu 18.04.

\subsection{Experimental Results}

\subsubsection{RQ1: Comparison with Baselines}
RQ1 aims to investigate whether {\sc deGraphCS} outperforms the state-of-the-art deep code search models. We evaluate {\sc deGraphCS} on the test set, which consists of 2,000 pairs of code snippets and the corresponding descriptions. In this automatic evaluation, we treat each description as an input query, and its corresponding code snippet as the ground truth. We report our evaluation results in Table I. The columns R@1, R@5, R@10 show the values of SuccessRate@k over all queries when $k$ is set to 1, 5 and 10, separately. The column MRR presents the MRR value of each model. From Table I, we can clearly find that {\sc deGraphCS} beats existing code search methods to a large extent on all the metrics. In concrete, {\sc deGraphCS} obtains an MRR of 51.73\%, which is much better than that of DeepCS (32.68\%), UNIF (41.93\%), MMAN (37.97\%). As for SuccessRate@k, {\sc deGraphCS} improves the state-of-the-art R@1 score from 34.05\% (obtained by MMAN) to 43.05\%. For 43.05\%/61.75\%/68.10\% of the test queries, the relevant code snippets can be found within the top 1/5/10 returned results by {\sc deGraphCS}. The results confirm that the improvement achieved by {\sc deGraphCS} is statistically and consistently significant, which indicates the effectiveness of {\sc deGraphCS}.

\begin{table}[t]
    \caption{Comparison of the overall performance between our model and baselines on automatic evaluation metrics}
    \centering
    \begin{tabular}{lcccc}
            \toprule
            Method & R@1 & R@5 & R@10 & MRR\\
            \midrule
            DeepCS & 0.2350 & 0.4185 & 0.5045 & 0.3268\\
            UNIF & 0.3250 & 0.5175 & 0.5980 & 0.4193 \\
            MMAN & 0.3405 & 0.5325 & 0.6130 & 0.4342 \\
            deGraphCS & \textbf{0.4305} &\textbf{0.6175} & \textbf{0.6810} & \textbf{0.5173}\\
            \bottomrule
        \end{tabular}    
        \label{tab1}
\end{table}

\subsubsection{RQ2: Effects of Integration}
To answer this question, we perform experiments over different combination mechanism of the features (\ie code tokens, variable-based data and control graph) by two models: {\sc deGraphCS} and MMAN (Token+V-DFG+V-CFG). Table II shows the overall performance of the two approaches.

From Table II, we can observe that the MRR of MMAN (Token+V-DFG+V-CFG) decreases by 8.96\% compared with {\sc deGraphCS}. In terms of SuccessRate@k, MMAN (Token+V-DFG+V-CFG) achieves a SuccessRate@1/5/10 of 33.80\%, 52.50\% and 59.90\% respectively, much lower than those of {\sc deGraphCS}. It means that more relevant code snippets can be returned by {\sc deGraphCS}. Therefore, integrating the three features into one graph preforms better than roughly fusing them by a single attention layer.

\begin{table}[t]
     \caption{Effect of Graph Integration} \centering
    
    \setlength{\tabcolsep}{0.99mm}{
        \begin{tabular}{lcccc}
            \toprule
            Method & R@1 & R@5 & R@10 & MRR\\
            \midrule
            MMAN(Token+V-DFG+V-CFG) & 0.3380 & 0.5250 & 0.5990 & 0.4277 \\
            deGraphCS & \textbf{0.4305} & \textbf{0.6175} & 
            \textbf{0.6810} & \textbf{0.5173} \\
            \bottomrule
        \end{tabular}   } 
        \label{tab2}
  
\end{table}

\subsubsection{RQ3: Effect of Graph Optimization Component}
To demonstrate the effectiveness of the flow graph optimizing mechanism constructed from LLVM IR, we perfrom experiments by comparing {\sc deGraphCS} with the version removing the graph optimization mechanism {\sc deGraphCS}-noGO. We present the overall comparison results in Table III. 

From Table III, we can see that {\sc deGraphCS}-noGO achieves average success rate of 36.20\%, 55.85\%, 62.10\% when the top 1, 5, 10 results are inspected, respectively, while {\sc deGraphCS} achieves 18.92\%, 10.56\% and 9.66\% improvement in terms of SuccessRate@1, SuccessRate@5 and SuccessRate@10, respectively. In terms of MRR, {\sc deGraphCS} achieves a 13.77\% improvement compared with {\sc deGraphCS}-noGO.  

The results demonstrate that removing redundant information and optimizing node contents of the initial constructed flow graph will make our proposed model focus more on the useful and fine-grained correlations between source code and the comment descriptions. Besides, the performance of {\sc deGraphCS}-noGO still outperforms state-of-the-art models, which further verifies the robustness of our proposed code representation method. Besides, we have also made some statistics and found that the total number of nodes in graph has been reduced by 51.88\%, which results in nearly half decreasing of training time.

\begin{table}[t]
   \caption{Effect of Graph Optimization} \centering
    \begin{center}
        \begin{tabular}{lcccc}
            \toprule
            Method & R@1 & R@5 & R@10 & MRR\\
            \midrule
            deGraphCS-noGO & 0.3620 & 0.5585 & 0.6210 & 0.4547 \\
            deGraphCS & \textbf{0.4305} &\textbf{0.6175} & \textbf{0.6810} & \textbf{0.5173}\\
            \bottomrule
        \end{tabular}     
        \label{tab3}
    \end{center}
\end{table}

\subsubsection{RQ4: Model Robustness}

\begin{figure}[t]
	\begin{minipage}[t]{0.49\textwidth}
		\centering
		\includegraphics[width=\textwidth]{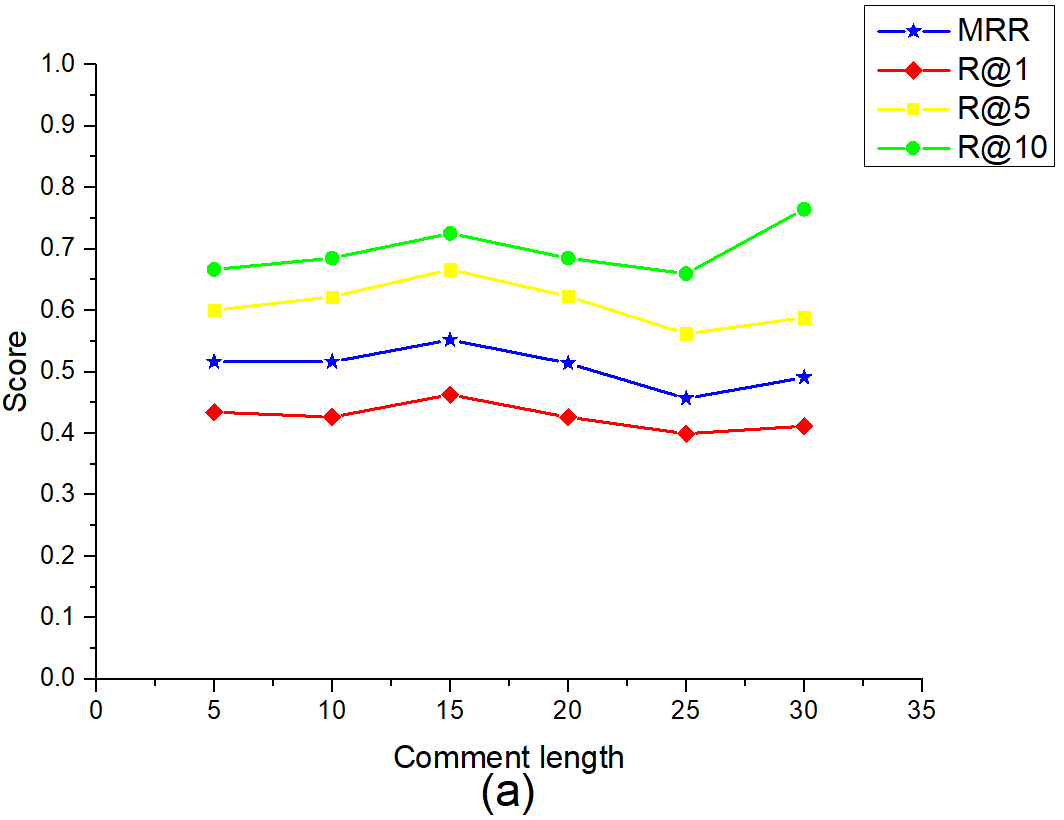}
	\end{minipage}
	\begin{minipage}[t]{0.49\textwidth}
		\centering
		\includegraphics[width=\textwidth]{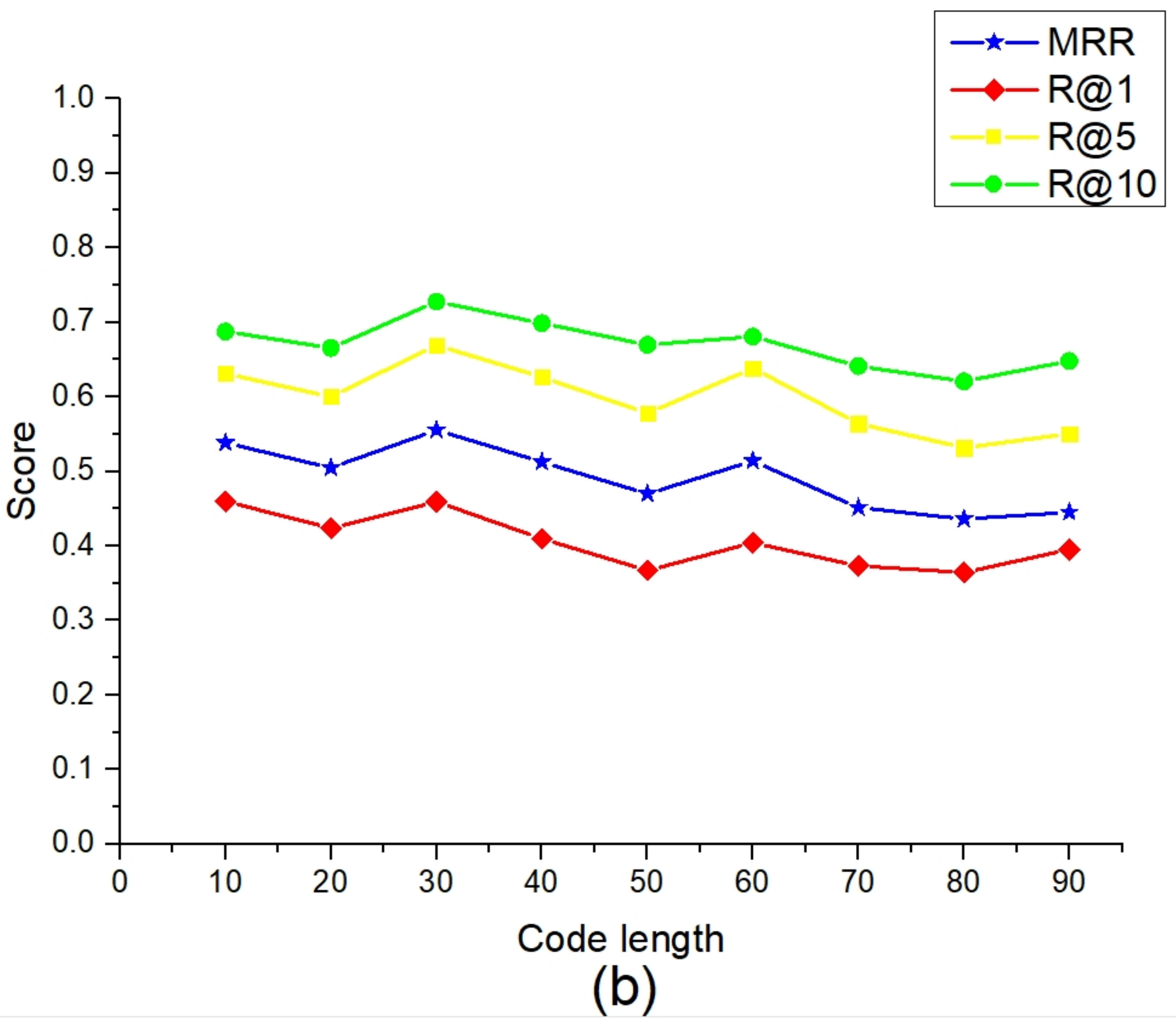}
	\end{minipage}
	\\
	\begin{minipage}[t]{0.49\textwidth}
		\centering
		\includegraphics[width=\textwidth]{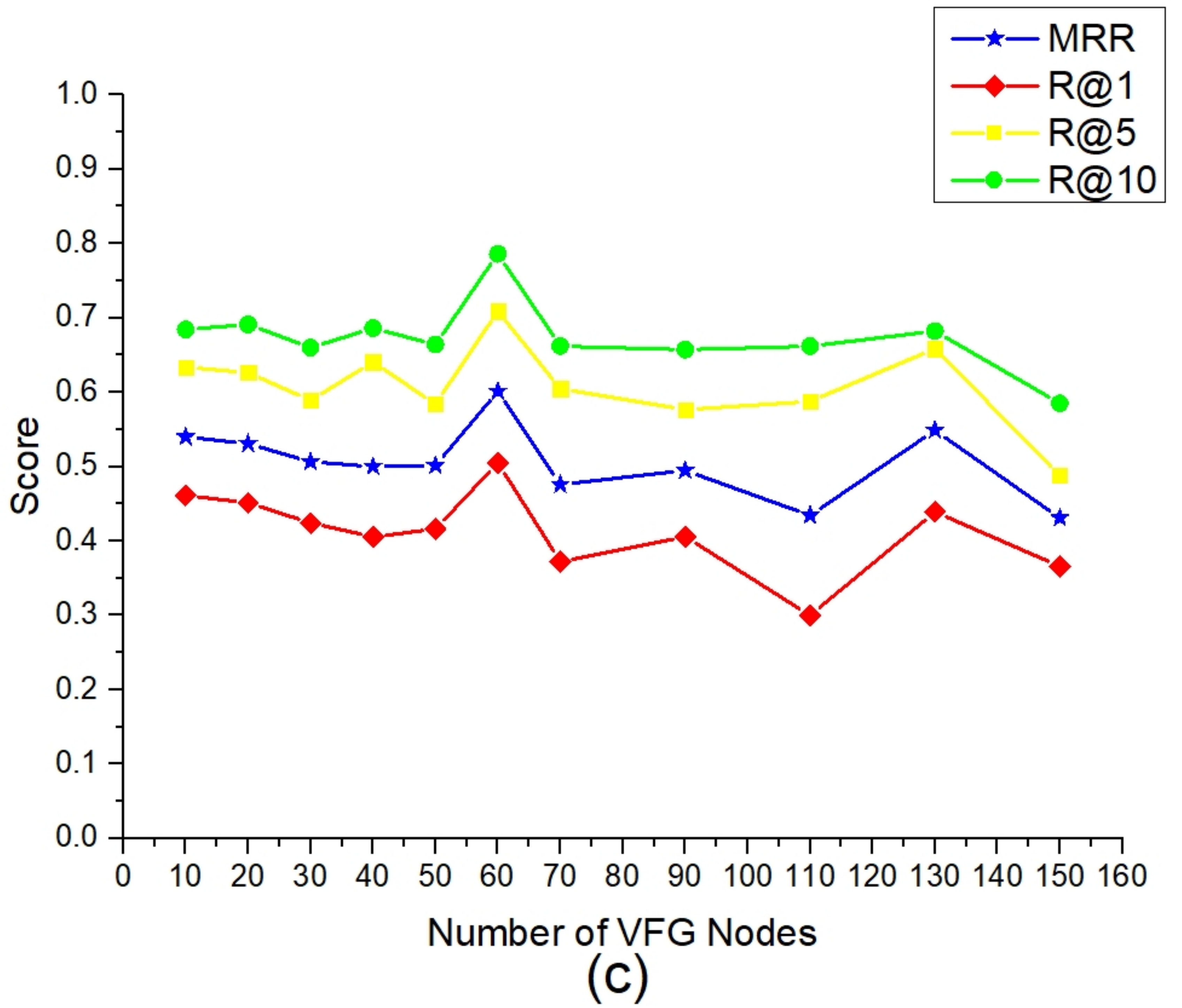}
	\end{minipage}
	\begin{minipage}[t]{0.49\textwidth}
		\centering
		\includegraphics[width=\textwidth]{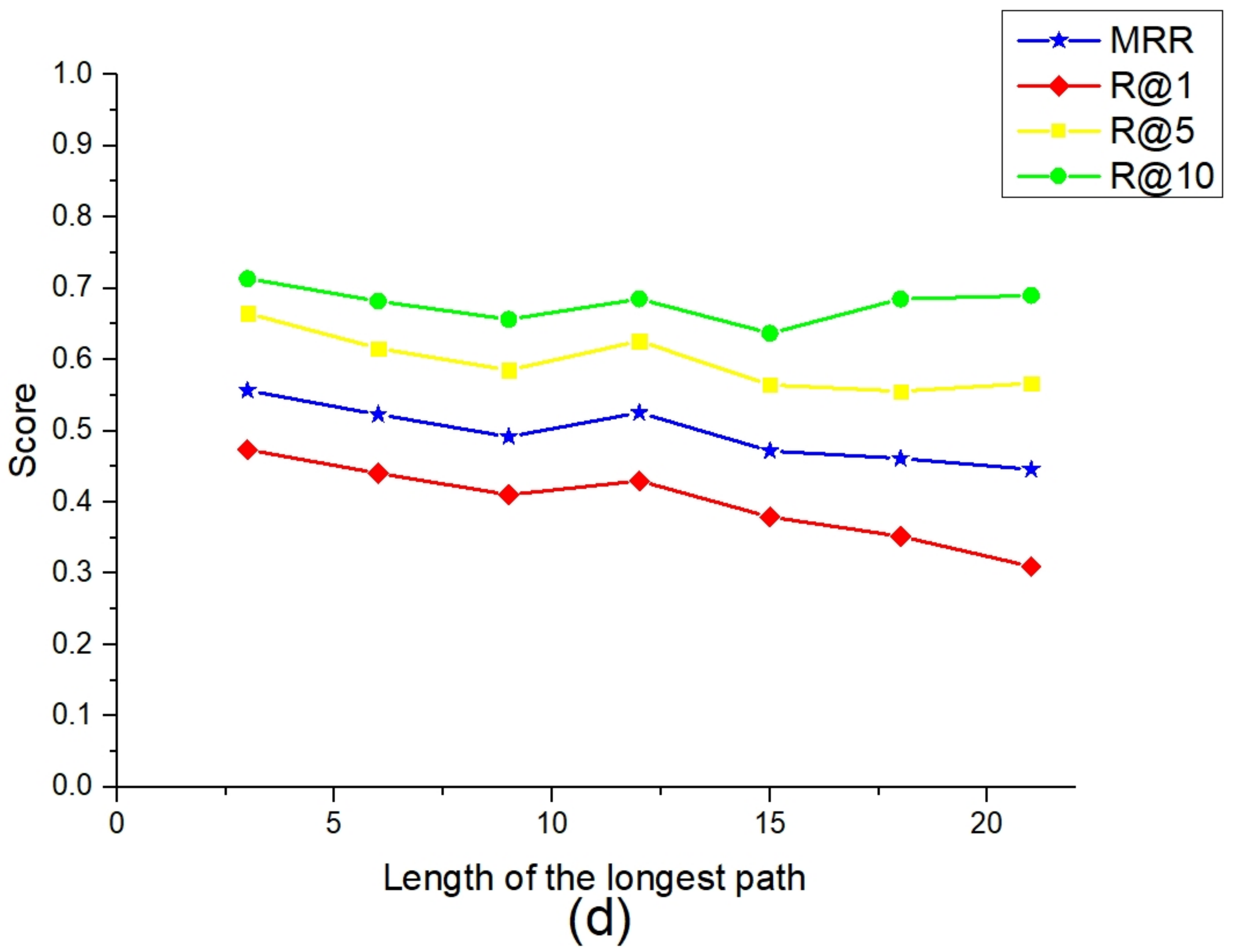}
	\end{minipage}
	\caption{Experimental results of deGraphCS on different metrics w.r.t. varying number of nodes and code lengths.}
\end{figure}

To analyze the sensitivity of {\sc deGraphCS}, we explore four parameters (\ie comment length, code length, number of VFG nodes, the length of the longest path in VFG) which may have an impact on the code and comment representation. Fig. 5 illustrate the performance of {\sc deGraphCS} on the basis of different evaluation metrics with varying parameters. From Fig. 5(a)(b)(c) we can find that in most cases, {\sc deGraphCS} maintains a stable performance even though the coment length, code length or node number increases dramatically, which can be attributed to the superiority of our variable-based flow graph. The length of longest path between any two nodes is used to measure the complexity of our flow graph. From Fig. 5(d), we can see that with the length of the longest path increases (\ie the difficulty of the graph embedding increases), the performance of {\sc deGraphCS} decreases slightly but overall maintains a stable level. Overall, the results in this subsection further verify the robustness of our proposed model.

\subsubsection{RQ5: Human Evaluation}
{\sc deGraphCS} has shown great utility in the aforementioned automatic evaluation experiments. However, in reality, the questions of how useful a returned code snippet is are likely best answered by human programmers. Thus, we conduct a user study where 5 experienced programmers are asked to grade the utility of code fragments returned by the baseline methods, \ie DeepCS, UNIF, MMAN and our proposed model {\sc deGraphCS}. 

In concrete, we first build a search platform which includes a search codebase of 30,799 C language functions and 50 queries select from the test set as benchmark. By using different models, the platform will recommend top 10 code snippets to users according to the query. Then, we recruit 5 graduate students from our school, who have rich experience on C projects and are competent enough to perform the user study, to evaluate the effectiveness of these models.

To simulate real scenarios, we let 5 participants (denoted as P1 - P5) use the four models in turn, to search related code of 50 queries. All participants have over 2-years/5-projects C programming experience. During the evaluation, we make sure that participants did not know which model the results searched by. For each query, they need to inspect the top 10 results returned by each model and label those results they believe are relevant to the query. Table IV shows the overall performance achieved by each model in terms of SuccessRate@10 and MRR.

\begin{table}[t]
    \caption{Comparison of the overall performance between our model and baselines on manual evaluation (SuccessRate@10$|$MRR)}
    \centering
    \setlength{\tabcolsep}{1.5mm}{
        \begin{tabular}{lcccccc}
            \toprule
            Method & P1 & P2 & P3 & P4& P5 & Aveg.\\
            \midrule
            DeepCS & 0.40$|$0.21 & 0.34$|$0.18 & 0.34$|$0.17 & 0.52$|$0.36 & 0.48$|$0.35 &0.42$|$0.26\\
            UNIF & 0.52$|$0.36 & 0.52$|$0.27 & 0.48$|$0.29 & 0.60$|$0.39 & 0.58$|$0.39 & 0.54$|$0.34\\
            MMAN & 0.58$|$0.41 & 0.52$|$0.33 & 0.48$|$\textbf{0.37} & 0.64$|$0.43 & 0.64$|$0.44 & 0.57$|$0.40\\
            deGraphCS & \textbf{0.66$|$0.47} &\textbf{0.62$|$0.46} & \textbf{0.56}$|$0.36 & \textbf{0.70$|$0.51} & \textbf{0.70$|$0.61} & \textbf{0.65$|$0.48}\\
            \bottomrule
        \end{tabular} } 
        \label{tab1}
\end{table}

From Table IV, we can draw the following conclusions: (a) under the experimental setting, {\sc deGraphCS} answer more user queries with an average SuccessRate@10 of 0.65 and the improvements compared with MMAN, UNIF and DeepCS is 14\%, 20\% and 55\%, respectively; (b) {\sc deGraphCS} achieves a better code search performance with an average MRR of 0.48. In conclusion, our proposed model maintains higher practice value in simulated code search scenario.

We further choose several examples to illustrate the superiority of {\sc deGraphCS} on the searching results. Fig. 6 shows the searched results of {\sc deGraphCS}, DeepCS, UNIF and MMAN on the query ``allocate memory for the file descriptors''. We can see that the result returned by {\sc deGraphCS} is exactly what the users need. However, the results returned by DeepCS, UNIF and MMAN do not realize the queried function and only focus on the shallow information (\ie keyword ``fd'', which is related to file descriptor). In concrete, the core functionality related to the keywords in the query is squared in red. The baseline methods can only retrieve the functions that realize the file creation, open or memory allocation for other data structures instead of file descriptor. Fig. 7 shows the rank 1 and rank 2 searched results of {\sc deGraphCS} on the query ``calculate checksum of checkpoint''. We can see that both the two retrieved code snippets realize the function of buffer checking and computing. However, compared with the rank 1 result, whose tokens are well-named (\ie checksum, buffer) and obviously matched with the keywords in the query, the variables in rank 2 code snippet are much more obscure (\ie cksum, b). Since the tokens in rank 2 code are not named following natural languages rules, it is hard for the models to utilize the token information for matching. The fact that {\sc deGraphCS} can retrieve the obscure code snippet related to the query on the semantics demonstrates that the data and control flow features are essential in code search and {\sc deGraphCS} can fully exploit the useful information in our variable-based flow graph. 

\begin{figure}[!t]
	\centering
	\includegraphics[width=0.8\textwidth]{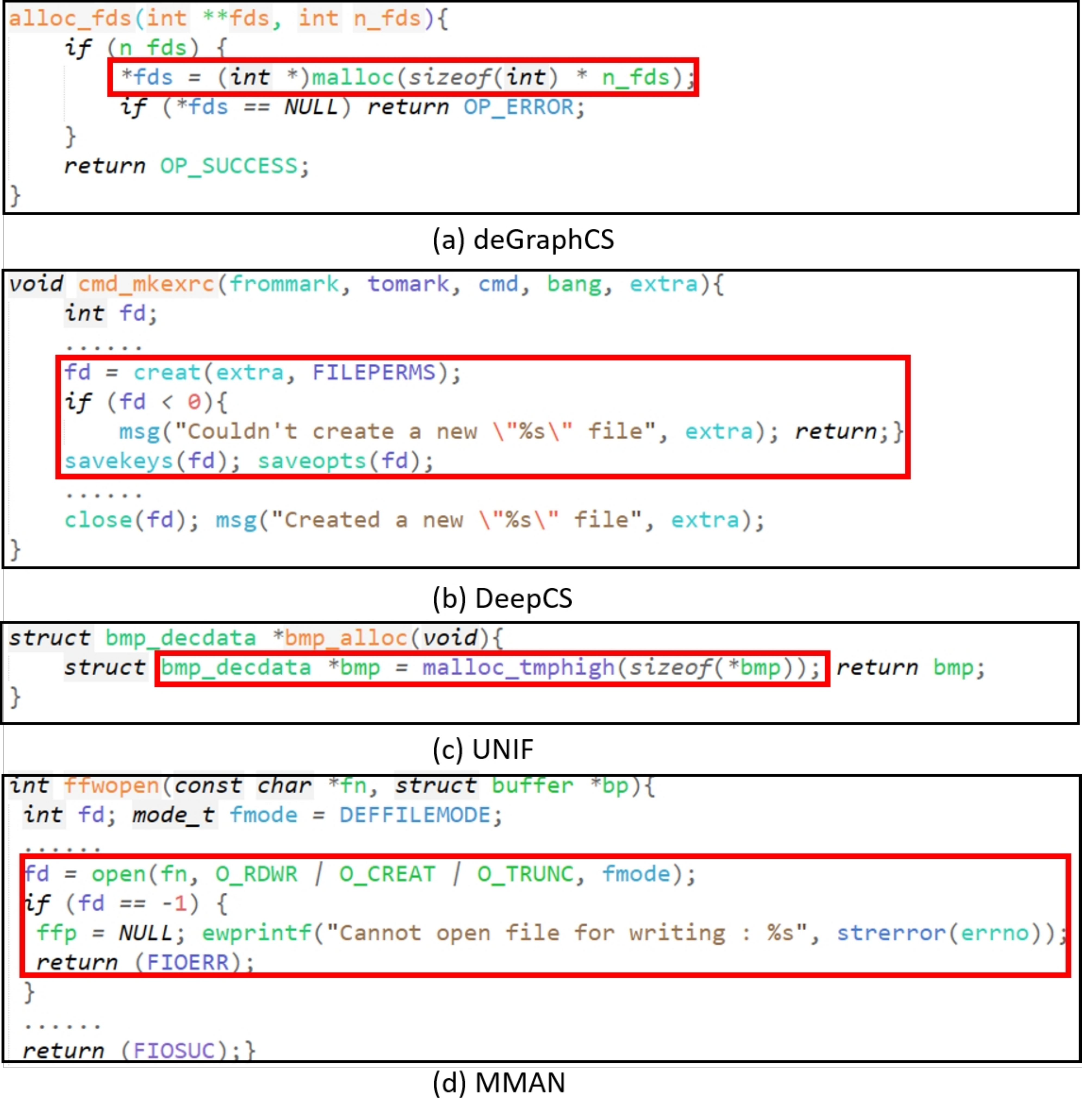}
	\caption{An illustrative example shows the comparison between the searched results of (a) deGraphCS, (b) DeepCS, (c) UNIF and (d) MMAN on the query ``allocate memory for the file descriptors''.}
	\label{fig1}
\end{figure}

\begin{figure}[!t]
	\centering
	\includegraphics[width=0.8\textwidth]{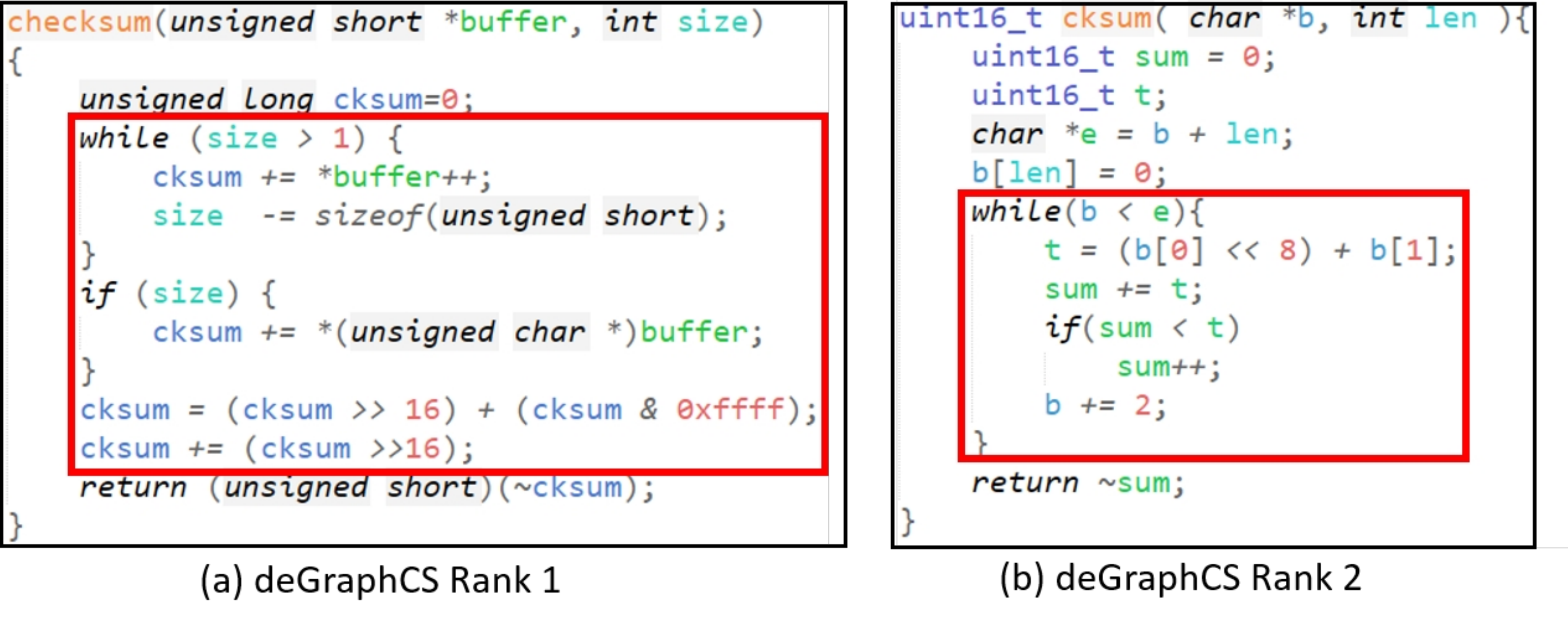}
	\caption{The rank 1 and rank 2 searched results of deGraphCS on the query ``calculate checksum of checkpoint''.}
	\label{fig1}
\end{figure}


\subsection{Threats to Validity}
{\sc deGraphCS} may suffer from three threats to validity. The first one lies in the scalability of our proposed approach. The LLVM IR can only be extracted from a whole program with complete dependencies. Therefore, it is difficult to extend our precise code representation model to some places where LLVM IR cannot be successfully extracted, such as many single code snippets in Stack Overflow. In fact, the difficulty can be solved by automatically adding the interface for those missing class objects and methods. And we plan to overcome these compilation problem to generate more dataset in the near future. The second threat is that {\sc deGraphCS} is currently trained and tested only on C programs. However, our work is based on LLVM IR of the code, which is independent of the source programming language. Thus, our code representation method can be easily transferred to other languages like Python and Java \etc For example, to transfer our work to Java language, we can use Soot\cite{Soot} (a Java optimization framework) to generate jimple which is intermediate representation of Java. And to transfer to python language, we can analysis and utilize the bytecode of python. We can build variable-based flow graph based on the above intermediate representation. The third threat lies in the model generalization ability. We construct a test dataset only consisting of 2,000 C code snippets, which may not be enough to represent most programming tasks. We plan to extend the dataset in the near future.

\section{Related Work}
\label{related_work}

\subsection {Code Search}
As vast repositories of open source code are available, a lot of works are proposed to search code to help developers. There are many works that try to search code according to user query. The traditional code search methods are mainly based on information retrieval and natural language processing technologies \cite{chan2012searching,holmes2009end,keivanloo2014spotting,li2016relationship,lv2015codehow,mcmillan2011exemplar,mcmillan2011portfolio,bajracharya2006sourcerer}. The methods mainly take text and structure characteristics of source code into consideration. \cite{bajracharya2006sourcerer} proposes a code search engine Sourcerer which extracts fine-grained structural information from source code. \cite{lv2015codehow} proposes CodeHow, a code search technique which reveals APIs related to user queries according to text similarity, and then applys an extended boolean model to utilize API information in code searching. NCS\cite{sachdev2018retrieval} utilizes natural language processing technologies to embed the code and query into vector, and then search the code snippet by comparing similarity of the vector representations. The above information retrieval-based methods treat both source code and query as natural language. It is hard to deeply understand the semantic information of source code.  

Since traditional code search based on syntactic are easy to return vague
or irrelevant result, many works\cite{semantic-codesearch2009,lightweight-spec,wang2014active,test-driven,codegenie} are proposed for semantic code search which is depends on specifications. For example, to search semantic related code, \cite{semantic-codesearch2009} proposed a architecture for semantics-based code search. The architecture utilize many different specifications which includes keywords, method signatures and test cases \etc The semantic search based on specifications performs well for finding relevant code but requires developers to write complex specifications. To reduce the requirement of specifications, \cite{lightweight-spec} proposes a new approach in which programmers only are required to specify lightweight, incomplete specifications which are in the form of input/output pairs and/or partial program fragment. Then the approach uses an SMT solver to automatically identify programs. While it also requires extra specifications to understand code semantics. Compared with the prior semantic code search, our method uses a static code-level analysis on source code, without the need to input extra specifications or run the code snippets. And our work focus on the scenario that users can simply use natural language to describe their intents. Last, we obtain code semantic by using a deep learning model to learn a representation of code instead of depending on the input and analysis of specifications.

Plenty of works are proposed to enrich information of the queries by refining including query expansion and reformulation\cite{haiduc2013automatic,hill2011improving,hill2014nl,lu2015query,wang2014active,dietrich2013learning,lemos2014thesaurus}. In concrete, \cite{haiduc2013automatic} trains a recommender (Refoqus) based on machine learning technologies, Refoqus can recommend a reformulation strategy according to the properties of queries. \cite{lu2015query} utilizes synonyms generated from WordNet to extend the queries. \cite{wang2014active} utilizes the feedback of users to reformulate queries.

Recently, to understand the deep semantics of the code and query, deep learning technologies have been applied to code search \cite{chen2018neural,defreez2018path,gu2018deep,gu2016deep,henkel2018code,huang2018api,wan2019multi,shuai2020improving,sachdev2018retrieval,cambronero2019deep}. \cite{chen2018neural} proposes BVAE including two Variational AutoEncoders (VAEs) to
model source code and natural language respectively, and jointly train two model to capture the closeness between the latent variables of the code and description. Many works measure semantic similarity of source code and query through joint embedding and deep learning technologies. CODEnn \cite{gu2018deep} extracts tokens, filenames and API sequences of code as the features, and embed these information and queries into a shared space so that code snippet can be retrieved by the vector of query. UNIF \cite{cambronero2019deep} extends NCS and further fine-tuned embedding of code and query by jointly deep learning. To utilize more information of source code, \cite{wan2019multi} proposed MMAN which use a multi-modal (tokens,AST and CFG) to represent source code. Similarly, we embed the source code and query into a common space to mine the semantic relationship. However, these works can not precisely represent the semantics of source code. 

\subsection{Code Representation}
Deep learning techniques on program analysis have attracted increasing attention. Many works focus on the representation of source code to perform a further software engineering research \cite{husain2018create,husain2018towards,white2016deep,wan2018improving,mou2014convolutional,piech2015learning,dam2016deep,wei2017supervised,raychev2014code,tufano2018deep}. \cite{raychev2014code} adopts a RNN and n-gram model for code completion. To capture the structure information of code, \cite{mou2014convolutional} proposes a novel novel tree-based convolutional neural network (TBCNN) to represent the ASTs of source code. \cite{wei2017supervised} proposes a framework (CDLH) which incorporates an AST-based LSTM to exploit the lexical and syntactical information. More recently, to extract more information, \cite{wan2019multi} proposes MMAN to combine multiple semantic information of code, which includes tokens, AST, and CFG of the source code. These researches focus on different representations of source code to capture the structural, syntactical and semantic information. However, these works can not represent the code precisely on semantics. Therefore, we pay more attention to the semantics of code, and propose our variable-based flow graph method. 

Several recent work\cite{code-similirity,NIPS2018Chen,dominator2020,IWSC2020} have tried to utilize intermediate representation to represent code. \cite{code-similirity} construct dependency graph to represent code based on instructions of Java bytecode. They use graph to record data and control dependencies between instructions, so that they can detect the similarity of code by use a subgraph isomorphism algorithm to analyze the similarity of dependency graph. \cite{NIPS2018Chen} also focuses on building a graph based on LLVM IR, to represent code. And they obtain well performance when they use skip-gram model to learn the graph representation and apply to code task. Different with these works, we aim to make a more precise code search, thus we construct a different graph based on variable instead of instructions, which ensures that the granularity of both comments and code representations in code search is consistent. And compared with the prior used method, we use a deep learning model (GGNN) to learn the semantic representation of code. Furthermore, to obtain a more precise representation of code, we make an optimization on the graph to decrease the noises and improve the training efficiency.

\section{Conclusion and Future Work}
\label{conculsion}
In this paper, we propose a deep graph neural network named {\sc deGraphCS} for code search. Instead of considering structural features of source code such as AST, {\sc deGraphCS} incorporates multiple semantic features, \ie tokens, variable-based data and control flow extracted from LLVM IR of code into flow graph. Furthermore, we put forward an optimization to remove the redundant information of the graph, followed by a gated graph neural network with attention mechanism to capture the critical information of code. In addition, we use a unified framework to learn the representation of natural language query and corresponding code snippet. We conduct several experiments on trained models, and the results of automatic evaluation and manual evaluation both demonstrate that our proposed approach is effective and outperforms the state-of-the-art approaches.

In future, we plan to investigate the performance of the {\sc deGraphCS} on other dataset of different programming languages, \eg Python and Java. We also plan to extend the variable-based flow graph we designed to solve other software engineering problems, \eg code translation\cite{NIPS2018Chen}, API recommendation \cite{huang2018api,Cai2019BIKER}, and code clone detection \cite{white2016deep}.

\balance
\bibliographystyle{unsrt}  
\bibliography{main.bib}

\end{document}